\documentstyle[pre,aps]{revtex}

\title{Dissipative Particle Dynamics with energy conservation}
\author{Pep Espa\~{n}ol\footnote{e-mail: pep@fisfun.uned.es}}
\address{Departamento de F\'{\i}sica Fundamental, UNED,
Apartado 60141, 28080 Madrid, Spain}

\begin{document}
\maketitle
\pacs{
\Pacs{47}{11$+$j}{Computational methods in fluid mechanics}
\Pacs{05}{70 Ln}{Non-equilibrium thermodynamics, irreversible processes}
      }
\begin{abstract}
Dissipative particle dynamics (DPD) does not conserve energy and this
precludes its use in the study of thermal processes in complex
fluids. We present here a generalization of DPD that incorporates an
internal energy and a temperature variable for each particle. The
dissipation induced by the dissipative forces between particles is
invested in raising the internal energy of the particles. Thermal
conduction occurs by means of (inverse) temperature differences. The
model can be viewed as a simplified solver of the fluctuating
hydrodynamic equations and opens up the possibility of studying
thermal processes in complex fluids with a mesoscopic simulation
technique. 
\end{abstract}

\section{Introduction}
Dissipative particle dynamics constitutes a valuable tool for
mesoscopic simulations of complex fluids. It was introduced by
Hoogerbrugge and Koelman \cite{hoo92},\cite{koe93} and has received
growing interest in view of its potentiality in the study of complex
flow problems as those arising in porous flow \cite{hoo92}, colloidal
suspensions \cite{koe93}-\cite{boe97b}, polymer suspensions
\cite{sch95}, or multicomponent flows \cite{cov97}. The technique has
received a considerable theoretical backup \cite{esp95}-\cite{mar97y}
that provides solid ground for its use.

DPD faces, however, a conceptual problem in that energy is not
conserved: the dissipative particles interact with dissipative forces
that depend on the relative velocities between particles. For this
reason, a DPD system cannot sustain a temperature gradient as it has
been pointed out by Marsh et al.\cite{mar97}. Nevertheless, in the
picture where the dissipative particles are understood as droplets or
mesoscopic clusters of atoms \cite{hoo92,esp97b} it is apparent that
the dissipated energy due to friction must be invested into increasing
the internal energy of the cluster. Therefore, we propose in this
letter to introduce an additional variable $\epsilon_i$ which is
interpreted as the internal energy of each particle. Along this new
variable we introduce an entropy variable $s_i=s(\epsilon_i)$ and a
temperature $T_i= [\partial s_i/\partial \epsilon_i]^{-1}$.

The variation of the internal energy must involve two different processes.
On one hand, temperature differences between particles produce
variations in the internal energies through ``heat conduction''. On the
other, the friction forces dissipate energy which is transformed into
internal energy through ``viscous heating''. Let us analyze each
process separately.

\section{Conduction}
Let us assume for a while that the $N$ dissipative particles of the
system are at rest at positions ${\bf r}_i$. We formulate the
following equation of motion for the internal energy of each particle

\begin{equation}
\dot{\epsilon}_i= \kappa \sum_j 
\left(\frac{1}{T_i}-\frac{1}{T_j}\right)\omega(r_{ij})+
\sum_j \tilde{q}_{ij}
\label{cond}
\end{equation}
This is a discrete fluctuating Fourier equation of heat conduction
with $\kappa$ playing the role of a thermal conductivity. We note that
the thermodynamic quantity conjugated to the internal energy is the
inverse of the temperature rather than the temperature
itself. $r_{ij}=|{\bf r}_i-{\bf r}_j|$ is the distance between
particles $i,j$ and the weight function $\omega(r)$ determines the
range of influence between particles. The random process
$\tilde{q}_{ij}$ mimics the random heat flux that occurs spontaneously
due to thermal fluctuations \cite{lan59}.  This random ``heat flux''
is assumed to be antisymmetric under particle interchange, this is,
$\tilde{q}_{ij}=-\tilde{q}_{ji}$ in such a way that the total internal
energy of the system is conserved $\frac{d}{dt}\sum_i \epsilon_i=0$ if
only a heat conduction process occurs.

The Langevin equations (\ref{cond}) have a mathematical well-defined
correspondence in the form of a stochastic differential equation (SDE),

\begin{equation}
d\epsilon_i= \kappa \sum_j 
\left(\frac{1}{T_i}-\frac{1}{T_j}\right)\omega(r_{ij})dt+
\sum_j \alpha \tilde{\omega}(r_{ij})d W^\epsilon_{ij}
\label{cond-sde}
\end{equation}
where we have expressed the random ``heat flux'' $\tilde{q}_{ij}$ from
particle $i$ to $j$ in terms of the increments of the Wienner process
$dW^\epsilon_{ij}$. The function $\tilde{\omega}(r_{ij})$ provides the
range of interaction and $\alpha$ is an overall noise amplitude. The
increments of the Wienner process are antisymmetric under particle
interchange $dW^\epsilon_{ij}=-dW^\epsilon_{ij}$ and satisfy the
mnemotechnical Ito rule

\begin{equation}
dW^\epsilon_{ii'}dW^\epsilon_{jj'}
=[\delta_{ij}\delta_{i'j'}-\delta_{ij'}\delta_{i'j}]dt
\label{ito}
\end{equation}
Following a standard procedure \cite{gar83} it is possible to obtain
the Fokker-Planck equation (FPE) that is mathematically identical to
the SDE (\ref{cond-sde}). The result is

\begin{equation}
\partial_t \rho_t(\epsilon) =
-\kappa \sum_{ij}\omega(r_{ij})
\frac{\partial}{\partial\epsilon_i}
\left[\frac{1}{T_i}-\frac{1}{T_j}\right]\rho_t(\epsilon)
+\frac{1}{2}
\sum_{ij}\alpha^2 \tilde{\omega}^2(r_{ij})\left[
\frac{\partial^2}{\partial\epsilon_i\partial\epsilon_i}-
\frac{\partial^2}{\partial\epsilon_i\partial\epsilon_j}\right]
\rho_t(\epsilon)
\label{fpe}
\end{equation}
where $\rho_t(\epsilon)$ is the density of probability of finding
a particular distribution $\epsilon_1,\ldots,\epsilon_N$ of internal
energies in each particle.

It is noteworthing that if one assumes the detailed balance condition
$\alpha^2 = 2\kappa$, and $\omega(r)=\tilde{\omega}^2(r)$, then the FPE
becomes

\begin{equation}
\partial_t \rho_t(\epsilon) = L^{HC}\rho_t(\epsilon)
\label{fpeok}
\end{equation}
where the heat conduction operator is given by
\begin{equation}
L^{HC}=
\kappa \sum_{ij}\omega(r_{ij})
\frac{\partial}{\partial\epsilon_i}
\left[\frac{1}{T_j}-\frac{1}{T_i} +
\left[
\frac{\partial}{\partial\epsilon_i}-\frac{\partial}{\partial\epsilon_j}
\right]\right]
\label{hc1}
\end{equation}
The FPE (\ref{fpeok}) admits the following unique equilibrium solution

\begin{equation}
\rho_{eq}(\epsilon) 
= \frac{1}{Z} \exp\left\{\sum_i s(\epsilon_i) \right\}
{\cal P}(\sum_k\epsilon_k)
\label{equilcon}
\end{equation}
where $Z$ is the normalizing factor and ${\cal P}(\sum_k\epsilon_k)$
is a function of the total internal energy and will be discussed in
the next section. This is the Einstein's formula for energy
fluctuations in the presence of energy conservation. Note that at
equilibrium the intensive parameters $T_i^{-1}$ are equal on average
\cite{esp89}

\begin{eqnarray}
\left\langle\frac{1}{T_i}\right\rangle 
&=&\int d\epsilon \frac{\partial}{\partial \epsilon_i}\left[\exp S(\epsilon)\right]
{\cal P}(\sum_k\epsilon_k)
\nonumber\\
&=& -\int d\epsilon \exp S(\epsilon)\frac{\partial}{\partial \epsilon_i}
{\cal P}(\sum_k\epsilon_k)
= -\int d\epsilon \exp S(\epsilon)\frac{\partial}{\partial \epsilon_j}
{\cal P}(\sum_k\epsilon_k)
=\left\langle\frac{1}{T_j}\right\rangle 
\label{eqint}
\end{eqnarray}
On the other hand, if one computes with the method of Lagrange
multipliers the most probable distribution
$\tilde{\epsilon}_i,\ldots,\tilde{\epsilon}_N$ subject to the
conservation of internal energy one also obtains
$T(\tilde{\epsilon}_i)=T(\tilde{\epsilon}_j)$ \cite{esp89}.

\section{Viscous heating} 
When the particles have certain velocities ${\bf v}_i$, the DPD model
prescribes the following equations of motion \cite{esp95}

\begin{eqnarray}
d{\bf r}_i&=& {\bf v}_i dt
\nonumber\\
d{\bf v}_i&=&\left[\sum_{j\neq i}\frac{1}{m}{\bf F}^C_{ij}({\bf r}_{ij})
-\sum_{j\neq i}\gamma_{ij}B(r_{ij})
({\bf e}_{ij}\!\cdot\!{\bf v}_{ij}){\bf e}_{ij}\right]dt
+\sum_{j\neq i}\sigma_{ij}\tilde{B}(r_{ij}){\bf e}_{ij}dW^v_{ij}
\label{motion}
\end{eqnarray}
where ${\bf F}^C_{ij}$ is a conservative force that derives from a
potential $V(r_{ij})$, ${\bf v}_{ij}={\bf v}_i-{\bf v}_j$ is the
relative velocity, and the functions $B(r),\tilde{B}(r)$ provide the
range of the forces. In the original DPD algorithm the friction
coefficient $\gamma_{ij}$ and the noise amplitude $\sigma_{ij}$ where
assumed to be identical for all pairs, this is, $\gamma_{ij}=\gamma$
and $\sigma_{ij}=\sigma$. We need to be more general now due to the
temperature variations. The increments of the Wienner process are
symmetric under particle interchange, $dW^v_{ij}=dW^v_{ji}$, in order
to ensure momentum conservation \cite{esp95}. They satisfy the Ito
mnemotechnical rule

\begin{equation}
dW^v_{ii'}dW^v_{jj'}
=[\delta_{ij}\delta_{i'j'}+\delta_{ij'}\delta_{i'j}]dt
\label{ito2}
\end{equation}

From the equation of motion (\ref{motion}) it is possible to evaluate
an infinitesimal increment of the total mechanical energy defined as
$E_{mec}=\frac{1}{2}\sum'_{ij}V(r_{ij}) + \sum_i\frac{1}{2} m v_i^2$.
The prime in the summatory excludes the terms $i=j$. Using Ito
calculus with the rules (\ref{ito2}) we obtain

\begin{equation}
dE_{mec} = - \frac{m}{2}\sum'_{ij}B(r_{ij})\gamma_{ij}
({\bf v}_{ij}\!\cdot\!{\bf e}_{ij})^2dt
+\frac{m}{2}\sum'_{ij}\tilde{B}^2(r_{ij})\sigma_{ij}^2dt
+\frac{m}{2}\sum'_{ij}\sigma_{ij}\tilde{B}(r_{ij})
({\bf v}_{ij}\!\cdot\!{\bf e}_{ij})dW^v_{ij}
\label{incene}
\end{equation}
We make the assumption that all the dissipated energy (\ref{incene})
is invested in rising the total internal energy
$\dot{E}=-\sum_i\dot{\epsilon}_i$ in such a way that the {\em total}
energy $E+\sum_i\epsilon_i$ is exactly conserved. This can be achieved
simply by considering the following SDE for the internal energy

\begin{equation}
d\epsilon_i 
=
\frac{m}{2}\left[\sum_{j}
\left[B(r_{ij})\gamma_{ij} ({\bf v}_{ij}\!\cdot\!{\bf e}_{ij})^2
-\sigma^2_{ij}\tilde{B}^2(r_{ij})\right]dt
-\sum_j\sigma_{ij}\tilde{B}(r_{ij})
({\bf v}_{ij}\!\cdot\!{\bf e}_{ij})dW^v_{ij}\right]
\label{visc}
\end{equation}

Now, if we assume that both processes of conduction and viscous
heating are operating then the proposed equations of motion are given
by

\begin{eqnarray}
d{\bf r}_i&=& {\bf v}_i dt
\nonumber\\
d{\bf v}_i&=&\left[\sum_{j\neq i}\frac{1}{m}{\bf F}^C_{ij}({\bf r}_{ij})
-\sum_{j\neq i}\gamma_{ij}B(r_{ij})
({\bf e}_{ij}\!\cdot\!{\bf v}_{ij}){\bf e}_{ij}\right]dt
+\sum_{j\neq i}\sigma_{ij}\tilde{B}(r_{ij}){\bf e}_{ij}dW^v_{ij}
\nonumber\\
d\epsilon_i&=& 
\frac{m}{2}\left[\sum_{j}
\left[B(r_{ij})\gamma_{ij} ({\bf v}_{ij}\!\cdot\!{\bf e}_{ij})^2
-\sigma^2_{ij}\tilde{B}^2(r_{ij})\right]dt
-\sum_j\sigma_{ij}\tilde{B}(r_{ij})
({\bf v}_{ij}\!\cdot\!{\bf e}_{ij})dW^v_{ij}\right]
\nonumber\\
&+& \kappa \sum_j \left(\frac{1}{T_i}-\frac{1}{T_j}\right)\omega(r_{ij})dt
+ \sum_j \alpha \tilde{\omega}(r_{ij})d W^\epsilon_{ij}
\label{dpdec}
\end{eqnarray}
We postulate that $dW^v_{ij}$ and $dW^\epsilon_{ij}$ are uncorrelated.

These stochastic differential equations have associated an equivalent
Fokker-Planck equation that governs the evolution of the distribution
function $\rho_t$ of the all the variables of the system. By assuming

\begin{eqnarray}
\tilde{\omega}^2(r)&=&\omega(r)
\nonumber \\
\alpha^2&=&2\kappa
\nonumber \\
\tilde{B}^2(r)&=&B(r)
\label{detbal}
\end{eqnarray}
and after some straightforward although lengthy
algebra the following FPE results

\begin{equation}
\partial_t \rho_t =L^{FP}\rho_t=\left[ L^C + L^{VH}+L^{HC}\right]\rho_t
\label{FPE}
\end{equation}
where $L^C$ is the usual Liouville operator of the conservative system
\begin{equation}
L_C
\equiv-\left[\sum_i{\bf v}_i\frac{\partial }{\partial {\bf r}_i}
+\sum'_{i,j}\frac{{\bf F}^C_{ij}}{m}\frac{\partial }{\partial {\bf v}_i}\right]
\end{equation}
The heat conduction operator $L^{HC}$ is given by (\ref{hc1}) and the viscous
heating operator $L^{VH}$ is given by

\begin{equation}
L^{VH}=
\frac{1}{2}\sum'_{ij}B(r_{ij})
L_{ij}\left[\gamma_{ij}({\bf v}_{ij}\!\cdot\!{\bf e}_{ij})
+L_{ij}\frac{\sigma^2_{ij}}{2}\right]
\label{vh1}
\end{equation}
where the operator $L_{ij}$ is defined by

\begin{equation}
L_{ij}={\bf e}_{ij}\!\cdot\!\left[
 \frac{\partial}{\partial {\bf v}_i}
-\frac{\partial}{\partial {\bf v}_j}
-\frac{m}{2}{\bf v}_{ij}
\left[
 \frac{\partial}{\partial \epsilon_i}
+\frac{\partial}{\partial \epsilon_j}
\right]
\right]
\label{op1}
\end{equation}

It is most important to realize that the Fokker-Planck operator in
(\ref{FPE}) satisfies $L^{FP}f(E)=0$ and $L^{FP}g(P)=0$ for any arbitrary
functions $f,g$ of the total energy and total momentum. This crucial
property is a reflection of the conservation of energy and momentum of
the SDE (\ref{dpdec}).

As in the previous section, the requirement that the FPE has a
prescribed equilibrium solution imposes certain
constrains on the different functions and parameters of the model.
We postulate that the equilibrium distribution function is given
by
\begin{equation}
\rho_{eq}(r,v,\epsilon)
=\frac{1}{Z}\exp\{\sum_i s(\epsilon_i)\} P(E_{tot}(r,v,\epsilon),{\bf P})
\label{equil}
\end{equation}
where $P(E,{\bf P})$ is a function of the total energy and momentum to
be determined below. It is easily shown that if $\sigma_{ij}=\sigma$, a
constant for all particles, and
\begin{equation}
\gamma_{ij}=\frac{m}{2}\sigma^2\frac{1}{2}\left[\frac{1}{T_i}+\frac{1}{T_j}\right]
\label{gamma}
\end{equation}
then the equilibrium distribution function (\ref{equil}) is the
solution of the FPE (\ref{fpeok}). This non-trivial result
(\ref{gamma}) constitutes along with (\ref{detbal}) the detailed
balance conditions for the model. The result (\ref{gamma}) is to be
compared with the detailed balance condition of the algorithm without
internal energy, $\gamma=m\sigma^2/2T$.

We discuss now the meaning of $P(E,{\bf P})$ by considering the
time-dependent probability density ${\cal P}(E,{\bf P})$ that the
system has a given value of the dynamical invariants $E,{\bf P}$. It
is given by

\begin{equation}
{\cal P}(E,{\bf P};t) = \int dz \delta(E-E(z) )\delta({\bf P}-\sum_k m{\bf v}_k)
\rho_t(z)
\label{dist}
\end{equation}
where $z$ is a shorthand for all the variables of the system and
$E(z)=\frac{1}{2}\sum'_{ij}V(r_{ij}) + \sum_i\frac{1}{2} m
v_i^2+\sum_i\epsilon_i$. The time derivative of (\ref{dist}) is zero,
$\partial_t {\cal P}=0$, as can be seen after using the Fokker-Planck
equation for $\partial_t\rho_t(z)$ and an integration by parts. This
means that the equations of motion do not modify the distribution of
dynamical invariants, which is, therefore, determined by the initial
distribution function $\rho_0(z)$. By using the equilibrium
distribution function $\rho_{eq}(z)$ in (\ref{dist}) one obtains
$P(E,{\bf P})={\cal P}(E,{\bf P})$ and the function $P(E,{\bf P})$
in (\ref{equil}) is
determined by the initial distribution of dynamical invariants.  A
similar argument can be given in classical mechanics where instead of
a Fokker-Planck operator a Liouville operator applies \cite{esp92}.  In
general, we will assume that $P(E,{\bf P})=P(E)\delta({\bf P})$
because the center of mass of the total system will be selected at
rest at the initial time.

A final comment on the meaning of the temperature is in order.
The rate of change of mechanical energy  is inferred
from equation (\ref{incene}),

\begin{equation}
\left\langle\dot{E}_{mec}\right\rangle = 
- \frac{m}{2}\left\langle\sum'_{ij}
\gamma_{ij}({\bf v}_{ij}\!\cdot\!{\bf e}_{ij})^2B_{ij}\right\rangle
+\frac{m\sigma^2}{2}\left\langle\sum'_{ij}B_{ij}\right\rangle
\label{rate}
\end{equation}
The first contribution is the energy dissipated by friction whereas
the second contribution is the energy provided by the noise. At
equilibrium both terms are equal and there is no net variation of the
mechanical energy, on average. These terms can be computed explicitly
with the distribution function (\ref{equil}) by using the invariance
of (\ref{equil}) under rotations (see appendix in
Ref. \cite{esp93}). It turns out then that the following relationship
is satisfied (for any particle $k$)

\begin{equation}
\left\langle\frac{\frac{1}{2}m{\bf v}_k^2}{\frac{D}{2}T(\epsilon_k)}
\sum'_{ij}B(r_{ij})\right\rangle=
\left\langle \sum'_{ij}B(r_{ij})\right\rangle
\label{equipartition}
\end{equation}
where $D$ is the physical dimension of space. This is a precise
statement about the relationship between the temperature of the
particle and its kinetic energy, at equilibrium. It constitutes the
equivalent to the equipartition theorem to which it reduces if all the
particles have a constant temperature and consistently the canonical
distribution is assumed.

In summary, we have presented an extension of DPD that includes an
internal energy variable. The mechanical energy dissipated by friction
is transformed into internal energy. In addition, heat conduction
occurs.  The macroscopic behavior of the system will be hydrodynamic
in the momentum and energy. It is possible to construct a kinetic
theory along the lines of Ref. \cite{mar97} for the DPD
model of this letter and obtain balance equations for the momentum and
energy. The energy now is a slow independent variable (because it is
conserved) and it is possible to have thermal gradients. This opens up
the possibility of study a large variety of problems in which thermal
processes in complex fluids are of concern.

Discussion with Marisol Ripoll during the elaboration of this work are
gratefully acknowledged. This work has been partially supported by a
DGICYT Project No PB94-0382 and by E.C. contract ERB-CHRXCT-940546.


\begin{thebibliography}{10}

\bibitem{hoo92} HOOGERBRUGGE P.J. and KOELMAN J.M.V.A.,
{\em Europhys. Lett.} {\bf 19}, 155 (1992).

\bibitem{koe93} KOELMAN J.M.V.A. and HOOGERBRUGGE  P.J.,
{\em Europhys. Lett.} {\bf 21}, 369 (1993).

\bibitem{boe97} BOEK E.S., COVENEY P.V., LEKKERKERKER H.N.W., and VAN
 DER SCHOOT P., {\em Phys. Rev. E} {\bf 55}, 3124 (1997). 

\bibitem{boe97b} BOEK E.S., COVENEY P.V. , and LEKKERKERKER H.N.W., {\em J. Phys.:
 Condens. Matter} {\bf 8} (in press) (1997).

\bibitem{sch95} SCHLIJPER A.G., HOOGERBRUGGE P.J., and MANKE C.W.,
{\em J. Rheol.} {\bf 39}, 567 (1995).

\bibitem{cov97}  COVENEY P.V.  and NOVIK K. E., {\em Phys. Rev. E}
{\bf 54}, 5134 (1996), {\bf 55}, 4831 (1997).

\bibitem{esp95}
ESPA{\~{N}}OL P. and WARREN P., {\em Europhys. Lett.} {\bf 30}, 191 (1995).

\bibitem{cov96} 
COVENEY P.V.  and ESPA\~NOL P., {\em J. Phys. A} {\bf 30}, 779 (1997).

\bibitem{esp96}
ESPA{\~{N}}OL P., {\em Phys. Rev. E}, {\bf 53}, 1572 (1996).

\bibitem{mar97} 
MARSH C., BACKX G., and ERNST M.H., {\em Europhys. Lett.}  {\bf 38},
411 (1997). MARSH C., BACKX G., and ERNST M.H., Phys. Rev. E (in
press).

\bibitem{mar97y} 
MARSH C.A. and YEOMANS J.M., {\em Europhys. lett.} {\bf 37}, 511 (1997).

\bibitem{esp97b} ESPA\~NOL P., SERRANO M., and Z\'U\~NIGA I., 
Int. J. Mod. Phys. C (1997) (in press).

\bibitem{lan59}LANDAU L.D. and LIFSHITZ E.M., {\em Fluid Mechanics}
(Pergamon Press, 1959)

\bibitem{gar83}
GARDINER C.W., {\em Handbook of Stochastic Methods},
(Springer Verlag, Berlin, 1983).

\bibitem{esp89} ESPA\~{N}OL J. and RUB\'I J.M., Physica {\bf A 161},
89 (1989).

\bibitem{esp92}	ESPA\~{N}OL J., F.J. DE LA RUBIA, and RUB\'I J.M.
	Physica {\bf A 187}, 589 (1992).

\bibitem{esp93} ESPA\~{N}OL P.  and   Z\'{U}\~{N}IGA I. , J. of
Chem. Phys., {\bf 98}, 574 (1993).


\end{thebibliography}
\end{document}